\documentclass[prd,showkeys,nofootinbib,preprint]{revtex4-1}

\usepackage{amsmath}
\usepackage{amssymb}
\usepackage{mathrsfs}
\usepackage{graphicx}
\usepackage{bbm}
\usepackage[]{hyperref}

\begin{document}
\title{de~Sitter invariance of the dS graviton vacuum}
\author{Atsushi Higuchi}
\email{atsushi.higuchi@york.ac.uk}
\affiliation{Department of Mathematics, University of York\\
Heslington, York, YO10 5DD, United Kingdom}
\author{Donald Marolf}
\email{marolf@physics.ucsb.edu}
\author{Ian A. Morrison}
\email{ian\_morrison@physics.ucsb.edu}
\affiliation{University of California at Santa Barbara,
Santa Barbara, CA 93106, USA}

\begin{abstract}The two-point function of linearized gravitons on de~Sitter space is infrared divergent in the standard transverse traceless synchronous gauge defined by $k=0$ cosmological coordinates (also called conformal or Poincar\'e coordinates).  We show that this divergence can be removed by adding a linearized diffeomorphism to each mode function; i.e., by an explicit change of gauge.  It follows that the graviton vacuum state is well-defined and de~Sitter invariant in agreement with various earlier arguments.
\end{abstract}


\maketitle


\section{Introduction}

Perturbative quantum gravity on inflating spacetime backgrounds is of substantial interest for many reasons.  At the phenomenological level, linearized gravitons induce tensor fluctuations in the cosmic microwave background (see e.g. \cite{Weinberg:2008zzc}) as well as (small) non-Gaussianities \cite{Maldacena:2002vr,*Maldacena:2011nz}. In addition, numerous authors have suggested that infrared (IR) effects associated with de~Sitter gravitons could lead to decay of the (effective) cosmological constant $\Lambda$ or surprisingly large loop corrections (see \cite{Seery:2010kh} for a recent review),
a breakdown of perturbation theory \cite{Giddings:2010nc,Giddings:2011zd},
or difficulties in defining gauge-invariant observables \cite{Giddings:2005id,*Giddings:2007nu}. Suggestions of decaying $\Lambda$ or large loop corrections stem from various well-known infrared divergences associated with gauge-fixed two-point functions $\langle h_{ab}(x) h_{cd}(y) \rangle$ in standard graviton vacuum states\footnote{Here we focus on pure Einstein-Hilbert gravity.  See \cite{Seery:2010kh} for a review of issues associated with massless scalars;   massive scalars with any $m >0$ do not cause correlation functions to diverge even at large arguments \cite{Marolf:2010zp,*Hollands:2010pr,*Marolf:2010nz,*Marolf:2011aa}.} falling into one of the following three (related) categories:  i) an explicit divergence at all $x,y$ from the sum over long-wavelength modes, ii) a divergence at large $x$ and $y$  (sometimes known as the ``growing variance"), or iii) a divergence at large relative separations between $x$ and $y$.

Our purpose here is to address the first divergence above involving the sum over long wavelength modes (though this will have implications for the 2nd issue as well).  As we review briefly below, this divergence arises \cite{Starobinsky:1979ty}  when the graviton two-point function on a de~Sitter background is evaluated in the natural de~Sitter-invariant state using the transverse traceless synchronous gauge
\begin{equation}
\label{TTS}
\nabla_a h^{ab} = 0, \ \ \  h^a_a =0, \ \ \  h_{\eta a} =0,
\end{equation}
associated with the conformally flat coordinates (also called $k=0$ cosmological coordinates or Poincar\'e coordinates) which cover half of de~Sitter space (see figure) and in which the $d$-dimensional de~Sitter metric is
\begin{equation}
\label{cc}
ds^2 = g^{\rm dS}_{ab} dx^a dx^b = \frac{\ell^2}{\eta^{2}} \left(-d\eta^2 + d\vec x^2 \right)
\end{equation}
for $d\vec x^2 = \sum_{i=1}^{d-1} (dx^i)^2$ and $\eta \in (-\infty, 0)$.  The spacetime index $a$ ranges over $\eta$ and the allowed values of the spatial index $i$.   The extent to which this divergence is physical has been of significant debate over the past 25 years or so.  For example, \cite{Allen:1986ta,Allen:1986dd,Allen:1986tt,Higuchi:2001uv} computed the two-point function in other gauges and found finite results.  In addition,
\cite{Allen:1986dd,Higuchi:1986py,Higuchi:2000ye} showed that the divergent part of the two-point function defined by (\ref{TTS},\ref{cc}) takes a form associated with linearized diffeomorphisms.   These references therefore conclude that a de~Sitter invariant vacuum exists.  This conclusion also follows from \cite{Higuchi:1991tn,Higuchi:1991tk}, which explicitly constructs a vacuum annihilated by the full set of
de~Sitter charges by working in transverse traceless synchronous gauge as defined by the so-called global coordinates on de~Sitter space (in which space at each time is a (compact) $(d-1)$-sphere so that no IR divergences can arise).  On the other hand \cite{Kleppe:1993fz,Miao:2009hb,Miao:2010vs,Miao:2011fc} argue that no de~Sitter-invariant vacuum can exist and raise various issues with the pro-invariance arguments given above (with the exception of that of \cite{Higuchi:1991tn,Higuchi:1991tk} to which we return in section \ref{disc}).  Though in this work we discuss de~Sitter invariance only at the linearized level, the reader should note that
de~Sitter invariance of the interacting graviton vacuum would prohibit any explicit decay of the cosmological constant (whose gradient would break de~Sitter invariance).  We also note that even linearized de~Sitter invariance implies that divergence (ii) discussed in the first paragraph is a gauge artifact (though it might still lead to non-local physical effects which manifest themselves differently in other gauges).

\begin{figure}
 \label{fig}
  \includegraphics[width=1.5in]{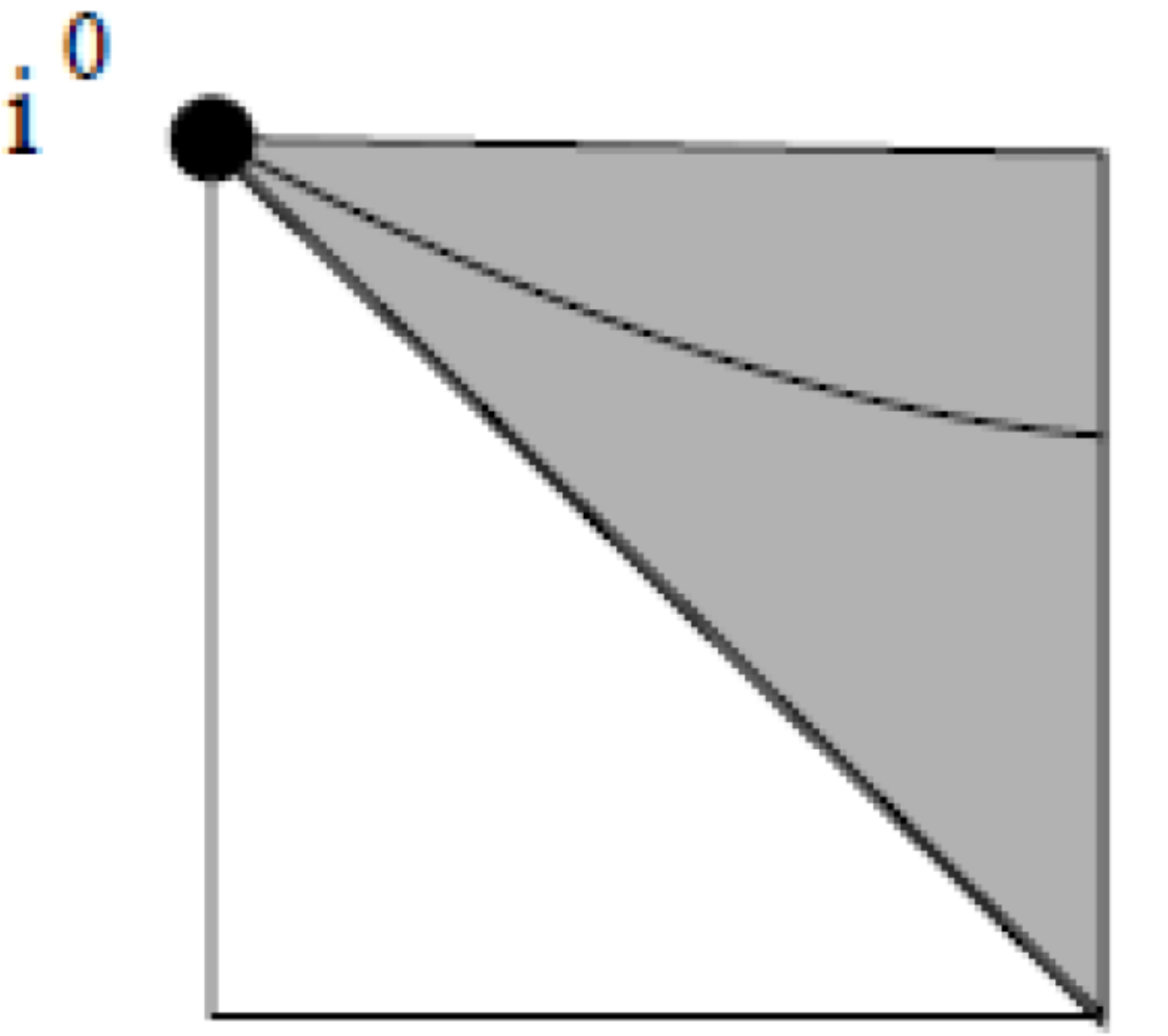}
  \caption{The region covered by conformally flat coordinates (\ref{cc}) is shown (shaded) on a conformal diagram of de~Sitter space.  Each point represents an $S^{d-2}$ which degenerates to zero size at the vertical lines at left and right. A surface of constant $\eta$ is also shown (curved line) which ends at the point $i^0$, corresponding to the region of large $\vec x$.
 }
\end{figure}

Below, we attempt to resolve this controversy by finding an explicitly allowed gauge transformation which renders the above two-point function finite over an arbitrarily large finite region of the $\vec x$ coordinates.  Our argument is closely related to the discussions of
\cite{Allen:1986ta,Unruh:1998ic,Higuchi:2000ye,Geshnizjani:2003cn,Lyth:2007jh,Bartolo:2007ti,Urakawa:2009my,Rajaraman:2010zx,Giddings:2010nc,Byrnes:2010yc,Urakawa:2010it,Gerstenlauer:2011ti,Xue:2011hm,Chialva:2011kx,Riotto:2011sf,Giddings:2011zd} in related contexts (and especially to \cite{Allen:1986ta,Higuchi:2000ye,Rajaraman:2010zx,Giddings:2011zd}) though our perspective is somewhat different.  Various subtleties involved with taking the region to be infinitely large and the detailed resolution we propose for the above controversy are presented in section \ref{disc}.

\section{A finite 2-pt function for the naive $k=0$ vacuum}
\label{main}

It is natural to study linearized gravity using the gauge defined by (\ref{TTS}, \ref{cc}) as the mode functions take a particularly simple form. The gauge conditions (\ref{TTS}) reduce the graviton equations of motion to those of a massless minimally-coupled scalar field \cite{Ford:1977dj} and translation invariance in $\vec x$ then allows solutions to be written in terms of plane waves.

For $d \ge 4$ the positive-frequency graviton mode functions with
definite wave number $\vec{k}$ are given by
\begin{equation}
\label{modes}
  \gamma^s_{ij}(\vec{k}) = \ell^{(6-d)/2} N
  \epsilon_{ij}^s(\vec{k})
  \eta^{(d-5)/2} H_{(d-1)/2}^{(2)}(k \eta) e^{i\vec k \cdot \vec x}, \quad k = |\vec k|.
\end{equation}
To prevent confusion below, we have chosen to denote individual mode functions by $\gamma_{ab}$ while reserving the symbol $h_{ab}$ for the full field operator (which is related to a sum over such modes).   Due to the synchronous condition $h_{\eta a} = 0$,  we can denote the modes by $\gamma_{ij}$, where $i,j$ run only over the $\vec x$ coordinates.
In (\ref{modes}), $\epsilon^s_{ij}(\vec{k})$ are an appropriate set
of (real) polarization tensors (for some set of polarizations $s$) whose indices are
raised/lowered by the $(d-1)$-dimensional Euclidean metric $\delta_{ij}$
and which are normalized according to
$\delta^{ik}\delta^{jl}\epsilon^{s_1}_{ij}\epsilon^{s_2}_{kl}
=  \delta^{s_1 s_2}$.  Here  $H^{(2)}_{(d-1)/2}(z)$ is
the positive-frequency Hankel function \cite{Abramowitz:1972} .
For $d=4$ this Hankel function reduces to the familiar expression
$H^{(2)}_{3/2}(k\eta) = i (2/\pi)^{1/2} (k \eta)^{-3/2}
(1+i k \eta) e^{-i k \eta}$.
The normalization coefficient is
$N = \left[ 4 (2\pi)^{d-1}  / \pi \right]^{-1/2}$.

With this normalization the above mode functions are orthonormal with respect
to the inner product
\begin{equation}
\label{eip}
  \langle \gamma^{s_1}(\vec{k}_1), \gamma^{s_2}(\vec{k}_2) \rangle
  := \frac{- i\ell^{d-2}}{(-\eta)^{d-2}} \int_{\eta = const}  d^{d-1} \vec x  \gamma^{s_1}_i{}^j(\vec{k}_1)
  \overleftrightarrow{\frac{\partial}{\partial \eta}} \gamma^{*s_2\,i}{}_j(\vec{k}_2)
  = \delta^{s_1s_2}\delta^{(d-1)}(\vec{k}_1-\vec{k}_2),
\end{equation}
where $i,j$ run only over spatial coordinates $\vec x$.
Since our modes satisfy (\ref{TTS}), the inner product (\ref{eip}) agrees with
that defined by the symplectic structure of \cite{Wald:1999wa} which we take to be the more fundamental definition (to be used in more general gauges).

One can use the modes (\ref{modes}) to define a gauge-fixed graviton operator
\begin{equation}
\label{gfh}
h^{gf}_{ij}(\vec x, \eta) = \sum_{s} \int d^{d-1} \vec k \left[a^\dagger_s(\vec k) \gamma^{*s}_{ij}(\vec k) + a _s (\vec k)\gamma^s_{ij}(\vec k)\right]
\end{equation}
with
\begin{equation}
\label{crels}
[a_s(\vec k),a^\dagger_{s'}(\vec{k}')] = \delta_{ss'} \delta^{(d-1)}(\vec k - \vec k').
\end{equation}
As is well-known, the two-point function in the natural vacuum state (defined by $a_s(\vec k) |0 \rangle = 0$) diverges at small $k$ \cite{Starobinsky:1979ty}:
\begin{equation}
\langle 0 |h^{gf}_{ij}(\vec x,\eta) h^{gf}_{mn}(\vec x',\eta')| 0 \rangle = \sum_s \int d^{d-1}\vec{k} \gamma^s_{ij}(\vec k) \gamma^{s*}_{mn}(\vec k) \sim \int \frac{d^{d-1}\vec{k}}{k^{d-1}}  \sim \ln k.
\end{equation}
This state is often called the Bunch-Davies vacuum.

However, this divergence is easily removed through the action of a linearized diffeomorphism.  The vector fields
\begin{equation}
\label{lds}
\xi^\eta (s,\vec k) = 0 ,   \ \ \ \xi_i (s,\vec k) = \eta^{-2} A(s, \vec k) \epsilon^s_{ij}x^j
\end{equation}
(with constant $A(s, \vec k)$) obey
\begin{equation}
\label{xi}
\nabla_{(a} \xi_{b)} (s,\vec k) = 0 \ \ \ {\rm for} \ a = \eta, \ \ \
\nabla_{(i} \xi_{j)} (s,\vec k) = \eta^{-2} A(s, \vec k) \epsilon^s_{ij},
\end{equation}
where $\nabla_a$ is the covariant derivative defined by (\ref{cc}).
Using the modified modes
\begin{equation}
\label{tg}
\tilde \gamma^s_{ab}(\vec k) = \gamma^s_{ab}(\vec k) + \nabla_{(a} \xi_{b)}(s, \vec k),
\end{equation}
choosing  $A(s, \vec k) =
- \ell^{(6-d)/2}
  N \frac{i 2^{(d-1)/2}}{\pi}\Gamma\left(\frac{d-1}{2}\right)\frac{e^{-\rho k}}{k^{(d-1)/2}}$ for some $\rho > 0$, and using
  \begin{equation}
    \eta^{(d-5)/2} H^{(2)}_{(d-1)/2}(k \eta)
  = \frac{i 2^{(d-1)/2}}{\pi}\Gamma\left(\frac{d-1}{2}\right)
  \eta^{-2} k^{(1-d)/2} + O\left(k^{(5-d)/2}\right) ,
  \end{equation}
   yields
\begin{equation}
\label{tmodes}
\tilde \gamma^s_{ij}(\vec k) = O(k^{(3-d)/2})
\end{equation}
at small $k$.   The two-point function of $\tilde h^{gf}_{ij}$ (defined as in (\ref{gfh}) with $\gamma^s_{ij}(\vec k)$ replaced by $\tilde \gamma^s_{ij}(\vec k)$ without changing (\ref{crels})) is thus finite at all finite $\eta, \vec x$.  We have computed this correlator for $d=4$ (see the appendix).
For $\rho=0$ the result would be manifestly invariant under the dilatations defined by $(\eta, \vec x) \rightarrow (\alpha \eta, \alpha \vec x)$ for constant $\alpha$, though it would not be invariant under translations $\vec x \rightarrow \vec x + \vec x_0$, which would be equivalent to multiplying
each $\xi_i (s,\vec k)$ by a phase.  However, setting $\rho =0$ would also introduce a logarithmic ultra-violet divergence.  We therefore restrict to $\rho > 0$.

The reader will note that each $\xi^{a} (s,\vec k)$ above diverges linearly at large $\vec x$.  It is thus far from clear that $\nabla_{(a} \xi_{b)} (s,\vec k)$ is pure gauge, or even that it is an allowed addition.  An argument in favor of using $\tilde \gamma^s_{ij}$ is that direct computation shows $\langle \nabla_{(a} \xi_{b)} (s, \vec k), \gamma^{s'}_{ij}(\vec k') \rangle =0$ when smeared against any smooth function of $\vec k'$, so that in particular the $\tilde \gamma^s_{ij}$ also satisfy (\ref{eip}).  On the other hand, it is certainly natural to impose fall-off conditions that require both metric perturbations and gauge transformations $\xi^a$ to vanish at large $\vec x$ (forbidding the use of (\ref{lds})), and one wonders how allowing
our $\xi^{a} (s,\vec k)$ might affect the interacting theory.

To eliminate concern over the behavior of $\xi^{a} (s,\vec k)$ at large $x$, we modify the above proposal as follows.  Choose two compact sets $K_0, K_1 \subset \mathbb{R}^{d-1}$ with $K_1 \subset {\rm int}(K_0)$ (where int$(K_0)$ denotes the interior of $K_0$) and replace the $\xi^a$ of (\ref{lds}) with any smooth vector fields $\xi^{a,K} (s,\vec k)$ which agree with $\xi^a (s,\vec k)$ inside $K_1$ but vanish outside $K_0$.  Such $\xi^{a,K} (s,\vec k)$ must be pure gauge, and are clearly null directions of the symplectic structure of \cite{Wald:1999wa}.\footnote{\label{afoot} As a result, we have $a^\dagger_s( \vec k) := \langle \gamma^s_{ij}(\vec k), h_{ab} \rangle =  \langle \gamma^{s,K}_{ij}(\vec k), h_{ab} \rangle$, where we take the inner product $\langle,\rangle$ to be defined by the symplectic structure of \cite{Wald:1999wa}, $h_{ab}$ is the full graviton field operator without gauge fixing, and $\gamma^{s,K}_{ab}(\vec k) = \gamma^{s}_{ab}(\vec k) - 2 \nabla_{(a} \xi_{b)^K} (s,\vec k)$ .  I.e., the creation/annihilation operators $a_s( \vec k),a^\dagger_s( \vec k)$ defined in this way are independent of $\xi^{a,K} (s,\vec k)$ and, in this sense, gauge-invariant.}   Making the same replacement in (\ref{tg}) defines new modes $\gamma_{ab}^{s,K}$ which can be used as in (\ref{gfh}) to define a gauge-fixed graviton operator $h^{gf,K}_{ab}$.  This operator and all of its correlators are then well-defined and finite inside $K_1$.  We note that $K_1$ can be taken arbitrarily large and, furthermore, the value of the gauge-fixed two-point function within some given $K_1$ always agrees with that of $\tilde h^{gf}_{ab}$ (see appendix) and does not change under either enlargement of $K_1$ or modification of $\tilde \xi^a (s,\vec k)$ outside $K_1$.

\section{Discussion}
\label{disc}

We have shown that the divergence in the standard transverse traceless synchronous 2-point function defined by $k=0$ cosmological coordinates (\ref{cc}) can be removed in any compact spatial region $K_0 \subset {\mathbb R}^{d-1}$ by a gauge transformation.  As a result, the vacuum $|0\rangle$ is meaningful without any cut-off being imposed on the {\it state}.  A similar conclusion might be reached directly from the observation that the Weyl-tensor two-point function defined by (\ref{TTS},\ref{cc}) is finite and de~Sitter invariant.\footnote{This observation follows from the results of \cite{Allen:1986dd}.  The correlator should agree with those found in other gauges, see e.g. \cite{Kouris:2001hz,Faizal:2011iv}.}  However, if we impose natural boundary conditions that require any gauge transformation must vanish at large $\vec x$, the action of $|0\rangle$ on fields in any given gauge remains ill-defined at large $x$.   In particular, the two-point function $\langle 0 | h^{gf,K}_{ab}(x)  h^{gf,K}_{cd} (y) |0\rangle$ is infinite whenever $x,y$ lie outside $K_0$.  We gave a family of prescriptions which break all de~Sitter symmetries except rotations in $\vec x$, but for which the two-point function within $K_0$ may nevertheless be computed in closed form.

We have so far chosen $K_0$ to be independent of $\eta$ for simplicity, but there is no harm in allowing time dependence.  In particular, we may choose the size of $K_0$ to grow at the speed of light as
$\eta \rightarrow -\infty$, as this still requires $\xi^{a,K}(s, \vec k) =0$ at large $\vec x$ at any fixed $\eta$.  Such a choice may allow the use of our $h^{gf,K}$ two-point function as a propagator for in-in perturbation theory, which the reader should recall computes a given $n$-point function using integrals only over the past light cones of the arguments.  As a result, for such calculations one may in practice be able to use the $\tilde h^{gf}$ propagator (discussed in more detail in the appendix for $d=4$).  However, one must also study the effect of the instantaneous (and thus non-local) Coulomb-like interaction that arises in gauges like (\ref{TTS}) (and which is analogous to the instantaneous interaction of Maxwell theory in Coulomb gauge).

We expect the results of such computations to agree with those obtained by other methods, such as the graviton 3-point functions computed in \cite{Maldacena:2002vr,Maldacena:2011nz}.  In particular, we remind the reader that the creation and annihilation operators $a_s( \vec k),a^\dagger_s( \vec k)$ are gauge-invariant when defined by an appropriate inner product (see footnote \ref{afoot}).  Since {\it momentum}-space correlation functions (and, in particular, the power spectrum, bispectrum, etc) may be defined as vacuum expectation values of products of these operators, such correlators are also gauge invariant when defined in this way.  It is only the representation in position space which depends on a choice of gauge and which is subject to gauge-dependent divergences\footnote{This comment resolves an apparent conflict  between de~Sitter invariance and finiteness emphasized in \cite{Miao:2009hb}.}.

The state $|0\rangle$ is easily shown to be de~Sitter invariant in the sense that it is annihilated by all generators of the de~Sitter group $SO(d,1)$.  This is manifest for the subgroup $E(d-1) \times \mathbb{R}$ of $SO(d,1)$ which preserves the region (see figure) covered by the conformal coordinates (\ref{cc}) and which acts as the $(d-1)$-dimensional Euclidean group on $\vec x$ together with the scale transformations just mentioned.  But it is also true of the remaining generators of $SO(d,1)$, which may be called special conformal transformations.  To see this, one need only write the charges $\int_\Sigma \xi^a n^n T_{ab}$ (where $\xi^a$ is the associated Killing field of de~Sitter space, $n^b$ is the unit future-pointing normal to the hypersurface $\Sigma$, and $T_{ab}$ is the stress tensor of linearized gravitons) in terms of the creation/annihilation operators $a_s( \vec k),a^\dagger_s( \vec k)$ using normal-ordering with respect to the vacuum $|0 \rangle$.  Because (infinitesimal) special conformal transformations map positive frequency modes to positive frequency modes, all $a^\dagger a^\dagger$ and $aa$ terms must cancel.  This leaves only terms of the form $a^\dagger a$ which annihilate $|0\rangle$.  With this ordering prescription, computing the commutator of two charges in the quantum theory proceeds precisely as in the classical theory and thus gives the usual result.  We thus conclude that linearized gravitons have a de~Sitter-invariant vacuum state in agreement with \cite{Allen:1986ta,Higuchi:1986py,Higuchi:1991tn,Higuchi:1991tk}.

There are, however, a number of subtleties on which we should elaborate.  First, we emphasize that we have addressed only linearized gravitons.  Ref. \cite{Miao:2011fc} raises the interesting possibility that de~Sitter invariance may be broken at the interacting level due to some effect associated with the non-propagating Coulomb fields required to satisfy the gravitational constraints.  Such concerns are not addressed by our work above.

Second, although the state $|0\rangle$ is de~Sitter invariant, it admits no de~Sitter invariant graviton two-point function under the boundary conditions discussed above. Defining a graviton two-point function requires a choice of gauge and the requirement $\xi^a(s, \vec k) \rightarrow 0$ at large $\vec x$ (for fixed $\eta$) means that the two-point function at points $x, y$ always diverges for large enough $\vec x, \vec y$ even if the geodesic separation between $x,y$ is held fixed.

Third, there is no conflict between our results and the analysis of Kleppe \cite{Kleppe:1993fz} who showed that no propagator in \cite{Tsamis:1992xa} could be de~Sitter invariant, even up to gauge transformations.  The key point here is that \cite{Tsamis:1992xa} did not work in the exact gauge (\ref{TTS}) and thus did not study the state $|0\rangle$ above.  Indeed, since the propagators of \cite{Tsamis:1992xa} are finite for all finite $\eta, \vec x$, it is clear that they do not describe any state in the Hilbert space defined by our $|0 \rangle$. The difference, however, is a subtle one.  We expect that some propagator in the class described by \cite{Tsamis:1992xa} describes a state that differs from $|0\rangle$ only in the treatment of linearized diffeomorphisms that are non-vanishing at large $\vec x$.  We also anticipate that this propagator {\it is} de~Sitter-invariant up to the addition of linearized diffeomorphisms which are large at large $\vec x$.  Such large linearized diffeomorphisms (which we do not call gauge transformations) were implicitly excluded in \cite{Kleppe:1993fz} by the assumption of boundary conditions which eliminated residual gauge freedom (and thus any possibility of compensating gauge transformations associated with dilatations).  Similar comments apply to the propagator of \cite{Miao:2011fc}. One would like to verify these expectations in detail, but this is beyond the scope of the present work.

Fourth, we emphasize that while $h^{gf,K}_{ab}$ is useful on the Hilbert space associated with the ``Bunch-Davies" vacuum $|0\rangle$, it may not be useful on other spaces of graviton states.  For example, as noted in \cite{Miao:2011fc}, one may follow \cite{Vilenkin:1983xp} to define a graviton state on the patch (\ref{cc}) of dS in which correlators of the original $h^{gf}_{ab}$ (defined by $\gamma^s_{ab}$) are finite so that those defined in this state by our $h^{gf,K}_{ab}$ in fact diverge inside $K_1$.

Finally,
while both are de~Sitter invariant, it would not be correct to think of our vacuum $|0\rangle$ as precisely the same state as the vacuum $|0\rangle_{global}$ defined in \cite{Higuchi:1991tn,Higuchi:1991tk} using transverse traceless synchronous gauge in global coordinates.  The point is again that our $|0\rangle$ is defined only on field operators which approach (\ref{gfh}) at large $\vec x$ and thus for which two-point functions necessarily diverge outside some compact set\footnote{Interestingly, a careful analysis shows that the linearized diffeomorphisms needed to convert the global modes studied in \cite{Higuchi:1991tn,Higuchi:1991tk} into the gauge defined by (\ref{TTS},\ref{cc}) must diverge linearly at large $\vec x$ just as did the $\xi^a(s,\vec k)$ of section \ref{main}.}.  In contrast, two-point functions in $|0\rangle_{global}$ can be finite everywhere \cite{Higuchi:2002sc}.  Similarly, because $\vec x = \infty$ is not invariant under special conformal transformations, one would not expect these generators to be self-adjoint\footnote{Though they are symmetric.  The situation is analogous to that of $-i\partial_x$ on the half line acting on functions that vanish at $x=0$.} on any domain in the Hilbert space defined by $|0\rangle$, while they are clearly self-adjoint on the Hilbert space constructed in \cite{Higuchi:1991tn,Higuchi:1991tk}.

We have suggested above that technical issues associated with boundary conditions at large $\vec x$ are required to resolve conflicts in the literature.  However, we do not believe that they are relevant to physics as described by observers with finite resources.  Indeed, the fact that the point $i^0$ at large $\vec x$ (see figure) is causally disconnected from all points at finite $\vec x$ would seem to forbid this.   This physical idea is implied by \cite{Verch:1992eg} for linear Klein-Gordon fields, but should hold much more generally -- at least up to issues associated with the lack of local gauge-invariants in interacting quantum gravity.  We therefore expect the states $|0\rangle$ and $|0\rangle_{global}$ to be operationally indistinguishable for such observers.

Although our state $|0\rangle$ is de~Sitter invariant, this invariance is associated with many subtleties.  In contrast, the de~Sitter invariant vacuum $|0\rangle_{global}$ is more straightforward: Two-point functions in
$|0\rangle_{global}$ are either manifestly invariant under a given element of $SO(d,1)$ or else transform in a manner that can be compensated by a finite gauge transformation.  This suggests that, despite the apparent simplicity of plane waves over spherical harmonics, approaches based on global coordinates may provide more control over graviton calculations in de~Sitter than those based on (\ref{cc}).
We speculate that such an approach may lead to insight into other infrared issues as well, perhaps including the physics associated with the divergence of many two-point functions at large separations between their arguments.  Indeed, the fact that the propagator of \cite{Hawking:2000ee} contains no such divergence suggests that this is again a gauge artifact, though their propagator does contain other large IR effects.

{\bf Note Added:}

Shortly  after our paper was posted on arxiv.org,
ref.\ \cite{Miao:2011fc} appeared with claims that our procedure changes important physics.  In particular, \cite{Miao:2011fc} notes that replacing $h^{gf}_{ab}$ with either $\tilde h^{gf}_{ab}$ or $h^{gf,K}_{ab}$ changes the position-space commutation relations (so that they no longer vanish outside the light cone), a certain definition of the power spectrum, and various definitions of the `propagator' for the theory.    The calculations in \cite{Miao:2011fc} are correct, but concern manifestly gauge-dependent quantities.  In particular, we note that while the definition of the power spectrum used in \cite{Miao:2011fc} is a common one, it is clearly gauge dependent.  The gauge independent definition (which we use above) is in terms of correlators of the $a(\vec k), a^\dagger(\vec k)$.  We emphasize that, as a matter of principle, experiments can measure only gauge-independent quantities.  All such quantities (including commutators) are invariant under replacement of $h^{gf}_{ab}$ with $h^{gf,K}_{ab}$.  We therefore see no grounds for the suggestion of \cite{Miao:2011fc} that a corresponding replacement in QED would modify physical effects in the infrared.

\begin{acknowledgements}
DM would like to thank Steve Giddings, Don Page, Albert Roura and Richard Woodard for enlightening discussions on gravitons in de~Sitter space over many years.  He would also like to thank Shun-Pei Miao, Nick Tsamis, and Richard Woodard for detailed discussions of an early draft of this work.  This work was supported in part by the US National Science Foundation under grants PHY05-55669 and PHY08-55415 and by funds from the University of California.
\end{acknowledgements}

\appendix

\section{The explicit propagator for $\tilde h^{gf}_{ij}$ in $d=4$}

The IR-divergent (symmetrized version of the) two-point function $\langle 0 | h^{gf}_{ij}(\eta, \vec x)  h^{gf}_{mn}(\eta', \vec x') |0 \rangle $ for $d=4$ was originally computed in \cite{Allen:1986dd} (see also \cite{Higuchi:2000ye}). The IR-divergent part of the two-point function (i.e., the part proportional to $\psi_1$ in \cite{Allen:1986dd})  with the normalization of the field $h_{ab}$ in this paper (which differs by a factor of $\sqrt{2}$ from \cite{Higuchi:2000ye}) is

\begin{equation}
\Delta^{IR}_{ijmn}(\vec{x},\eta;\vec{x}',\eta')
= - \frac{1}{40\pi^2\ell^2}\log(\alpha^2\left[|\vec{x}-\vec{x}'|^2-(\eta-\eta')^2\right])
\theta^{(2)}_{ijmn}(\eta,\eta'),
\end{equation}
where the momentum is cut off at $|\vec{k}|=\alpha$,
and where
$\theta^{(2)}_{ijmn}(\eta, \eta') \equiv  P_{im}P_{jn} + P_{in}P_{jm} - \frac{2}{3}P_{ij}P_{mn}$
for $P_{im} = \frac{\ell^2}{\eta\eta'}\delta_{im}$ and similarly for $P_{ij}, P_{mn}$ (proportional to $\eta^{-2}$ and $\eta^{\prime -2}$ respectively).

It is then straightforward to compute the two-point function
$\langle 0 | h^{gf}_{ij}(\eta, \vec x)  h^{gf}_{mn}(\eta', \vec x') |0 \rangle $ by including the effect of the linearized diffeomorphism terms (\ref{xi}).  The result is that the infrared logarithm term $\Delta^{\rm IR}_{ijmn}$ above is replaced by

\begin{equation}
\tilde{\Delta}^{IR}_{ijmn}(\vec{x},\eta';\vec{x}',\eta') = - \frac{1}{40\pi^2\ell^2}
\log \frac{\rho^2\left[|\vec{x}-\vec{x}'|^2 - (\eta-\eta')^2\right]}
{\left[|\vec{x}|^2 - (\eta-i\rho)^2\right]\left[|\vec{x}'|^2 - (\eta'+i\rho)^2\right]}
\theta^{(2)}_{ijmn}(\eta,\eta').
\end{equation}

\addcontentsline{toc}{section}{Bibliography}
\bibliographystyle{JHEP}
\bibliography{./bibliography2}

\end{document}